\documentclass[reprint,prb,twocolumn,showpacs,superscriptaddress,aps,longbibliography]{revtex4-2}

\usepackage[colorlinks=true,citecolor=blue]{hyperref} 
\usepackage{graphicx}
\usepackage{bm}
\usepackage{amsmath}
\usepackage{amssymb}

\DeclareGraphicsExtensions{.png,.jpg,.eps}
\usepackage{xcolor}

\newcommand{\beq}{\begin{equation}}
\newcommand{\eeq}{\end{equation}}

\usepackage{mleftright} 

\newcommand{\figref}[1]{\mbox{Fig.~\ref{#1}}}

\newcommand{\appref}[1]{\mbox{Appendix~\ref{#1}}}
\renewcommand{\eqref}[1]{\mbox{Eq.~(\ref{#1})}}

\newcommand{\figpanel}[2]{Fig.~\hyperref[#1]{\ref*{#1}(#2)}}
\newcommand{\figpanels}[3]{Fig.~\hyperref[#1]{\ref*{#1}(#2)-(#3)}}
\newcommand{\figpanelNoPrefix}[2]{\hyperref[#1]{\ref*{#1}(#2)}}

\usepackage{siunitx}

\begin{document}

\author{Guangze Chen}
\affiliation{Department of Microtechnology and Nanoscience, Chalmers University of Technology, 41296 G\"{o}teborg, Sweden}

\author{Jose L. Lado}
\affiliation{Department of Applied Physics, Aalto University, 02150 Espoo, Finland}

\author{Fei Song}
\email{songfeiphys@gmail.com}
\affiliation{Kavli Institute for Theoretical Sciences, Chinese Academy of Sciences, Beijing 100190, China }

\title{Many-body Liouvillian dynamics with a non-Hermitian tensor-network kernel polynomial algorithm}


\begin{abstract}
Understanding the dynamics of open quantum many-body systems is a major problem in quantum matter. Specifically, efficiently solving the spectrum of the Liouvillian superoperator governing such dynamics remains a critical open challenge. Here, we put forward a method for solving the many-body Liouvillian spectrum and dynamics based on the non-Hermitian kernel polynomial method and tensor-network techniques. We demonstrate the faithfulness of our method by computing the dynamics of the dephasing quantum compass model with a gradient magnetic field and comparing it with exact results. In particular, we show that our method allows us to characterize the quantum Zeno crossover and the reduction of relaxation rate due to Stark localization in this model. We further demonstrate the ability of our method to go beyond exact results by exploring nearest-neighbor interaction effects on the Liouvillian dynamics, elucidating the interplay between Stark localization and many-body interactions. 
Our method provides an efficient solution to many-body Liouvillian spectrum and dynamics, establishing a methodology to explore large open quantum many-body systems.
\end{abstract}

\date{\today}

\maketitle

\section{Introduction}
Exploring quantum dynamics provides a versatile approach to characterize both closed and open quantum systems~\cite{Polkovnikov2011, sieberer2023universality}. Unlike the coherent dynamics in closed systems, which is solely generated by the Hamiltonian, open quantum dynamics is subject to both quantum coherence and dissipative effects. The interplay between coherent and dissipative dynamics can motivate open quantum systems to approach non-equilibrium steady states with exotic properties after a long-time evolution~\cite{Verstraete2009, diehl2010dissipation, bardyn2012majorana, nakagawa2021eta, Mi2024}. Moreover, beyond steady states, the open quantum dynamics shares deep connections with a variety of areas in quantum physics, including many-body physics~\cite{PhysRevLett.111.150403, PhysRevLett.109.020403, PhysRevE.92.042143,bouganne2020anomalous, PhysRevLett.124.130602, PhysRevLett.132.150401}, random matrix theory~\cite{PhysRevLett.124.100604, PhysRevB.105.L180201}, and non-Hermitian quantum mechanics~\cite{PhysRevLett.123.170401,PhysRevLett.127.070402, PhysRevB.103.085428, PhysRevLett.128.110402,2024arXiv240309569S,2024arXiv240610135S}. 

Under the Markovian assumption, the dynamics of an open quantum system follows the Lindblad master equation that is generated by a Liouvillian superoperator ~\cite{breuer2002theory, PhysRevA.98.042118}. Plenty of information can be extracted from the spectrum of the Liouvillian, such as the non-Hermitian topology that links to the presence of the Liouvillian skin effect~\cite{PhysRevLett.123.170401, PhysRevB.103.085428, PhysRevLett.127.070402, PhysRevA.78.042307}, and the complex level statistics as a criterion of the quantum chaos in open quantum systems~\cite{PhysRevLett.123.254101, PhysRevX.10.021019, PhysRevLett.127.170602}. Nevertheless, except for several specific models~\cite{prosen2008third, PhysRevB.99.174303, PhysRevLett.117.137202, PhysRevB.108.115127}, it is often very challenging to exactly solve the Liouvillian spectrum in many-body systems~\cite{RevModPhys.93.015008}. Thus,  efficient numerical methods are needed.

Here, we put forward a methodology to compute dynamical correlators of the Liouvillian superoperator,
that provide access to the Liouvillian spectrum and the Liouvillian dynamics,
by combining the non-Hermitian kernel polynomial method (NHKPM)~\cite{PhysRevLett.130.100401} with tensor-network-based techniques~\footnote{The source codes are available on \href{https://github.com/GUANGZECHEN/NHKPM.jl}{https://github.com/GUANGZECHEN/NHKPM.jl}}. The NHKPM is a generalization of the Kernel Polynomial Method~\cite{RevModPhys.78.275}, an efficient method to compute dynamical correlators in Hermitian systems~\cite{PhysRevB.90.115124,PhysRevResearch.1.033009}, to non-Hermitian cases. It relies on expanding the correlator into Chebyshev polynomials and recursively calculating the expansion coefficients. Due to the simple mathematical structure of the recursion relations, the NHKPM can be efficiently implemented with matrix-product states (MPS), making it suitable for many-body calculations with low-entangled states. For many-body Liouvillians, we find that, with a proper vectorization~\cite{PhysRevLett.93.207205} of the density-matrix such that it becomes a low-entangled state while transforming the Liouvillian into a non-Hermitian model with only short-range interactions, the NHKPM implemented with MPS can efficiently solve the Liouvillian spectrum.

We demonstrate our methodology with the dephasing quantum compass model~\cite{PhysRevB.99.174303} with a gradient magnetic field, known to reduce the relaxation rate of unequal-time correlations of spins due to Stark localization~\cite{RevModPhys.34.645, PhysRevLett.122.040606, morong2021observation,doi:10.1073/pnas.1819316116}. We first benchmark our method with exact results in small systems, where we demonstrate that our method faithfully computes the Liouvillian dynamics. This allows us to characterize quantum Zeno crossovers~\cite{PhysRevResearch.5.023204, PhysRevB.108.115127} and Stark localization effects in the dynamics. Furthermore, to eliminate finite-size effects that hinder Stark localization at small magnetic field gradients, we consider a large system size, where the faithfulness of our method is demonstrated based on the closed hierarchy of correlations~\cite{PhysRevB.109.L140302, barthel2022solving}. Finally, moving beyond the capability of exact methods, we show that our method can be applied to explore the influence of nearest-neighbor interactions on the Liouvillian dynamics.

\begin{figure}[t!]
\center
\includegraphics[width=\linewidth]{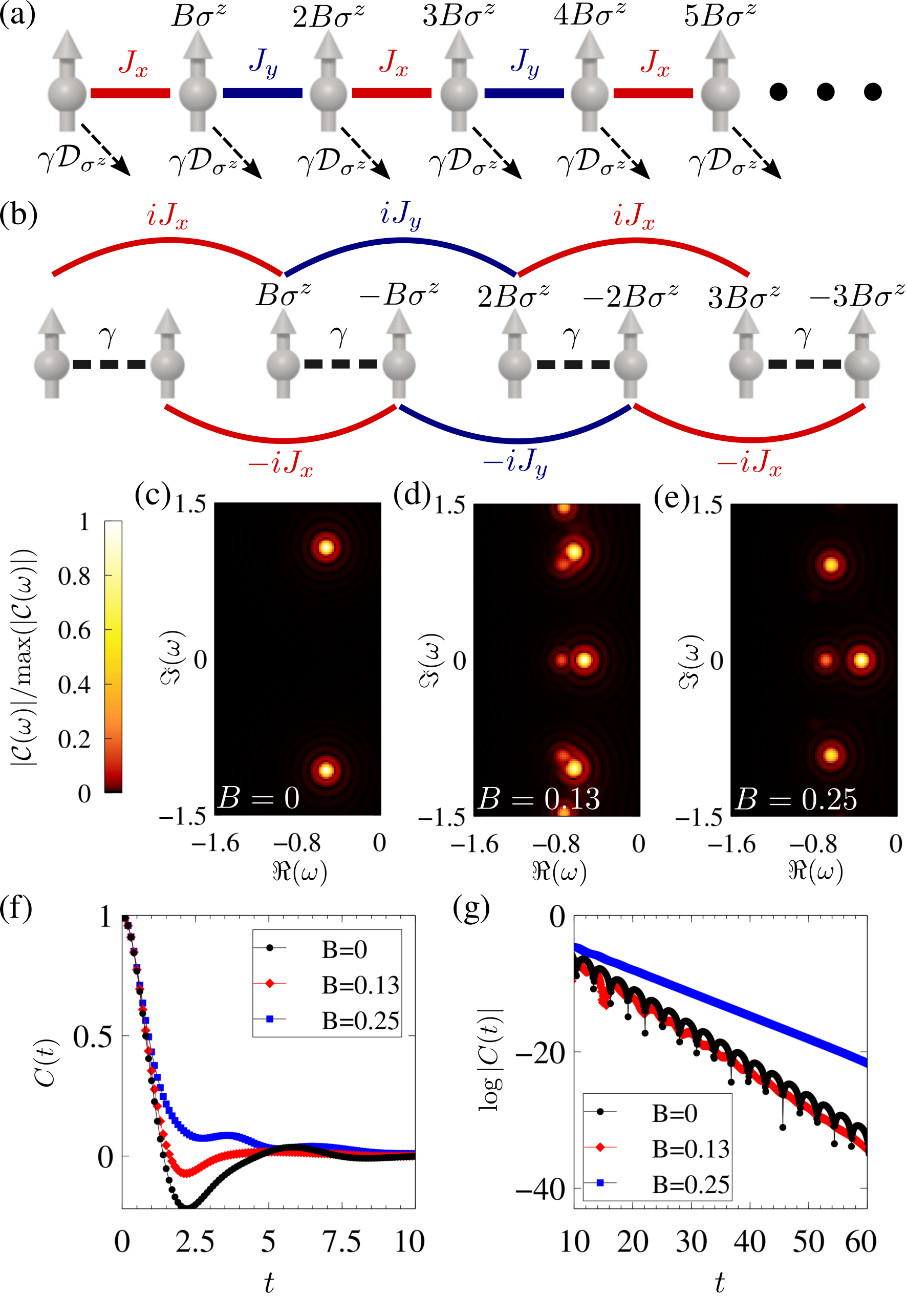}
\caption{The Liouvillian model and its dynamics. (a) The model Liouvillian~\eqref{eq_L} with $N$ spins, where $\mathcal{D}$ is the Lindblad dissipator [second term in~\eqref{eq_L}]. (b) The Liouvillian can be transformed into the non-Hermitian spin model~\eqref{eq5}. (c-e) The dynamical correlator of the Liouvillian $\mathcal{C}(\omega)$ in \eqref{eq9} computed with $N=4$, $J_x=0.75$, $J_y=0.5$, $\gamma=0.2$ and different magnetic field gradients $B=0,0.13$ and $0.25$. (f-g) The auto-correlator $C(t)$ in \eqref{eq7} computed with parameters in panels (c-e), showing the short and long-time dynamics. The long-time dynamics highlights the similar relaxation rates for $B=0$ and $B=0.13$, and a smaller relaxation rate for $B=0.25$.}
\label{fig1}
\end{figure}

\section{Model and Method}
We consider the Liouvillian 
\beq \label{eq_L}
\mathcal{L}[\rho]=-i[H,\rho]+\sum_{l}\left(L_l\rho L_l^\dag-\frac{1}{2}\{L_l^\dag L_l,\rho\}\right)
\eeq
with the following Hamiltonian:
\beq \label{eq1}
\begin{aligned}
H=-\sum_{l=1}^{N/2}J_x\sigma^x_{2l-1}\sigma_{2l}^x-\sum_{l=1}^{N/2-1}J_y\sigma^y_{2l}\sigma_{2l+1}^y+\sum_{l=1}^N B(l-1)\sigma_l^z
\end{aligned}
\eeq
and Lindblad dissipators $L_l=\sqrt{\gamma}\sigma^z_l$, where $l$ is the site index and $B$ is a gradient magnetic field [\figpanel{fig1}{a}]. The Lindblad dissipators describe the process of dephasing, which can originate from density-density coupling to a bath with infinite temperature such as a Floquet system~\cite{PhysRevA.82.063605}, or from successive measurements of $\sigma^z_i$~\cite{Preskill_note}. When the gradient magnetic field vanishes, i.e. $B=0$, this model reduces to the solvable quantum compass model with dephasing~\cite{PhysRevB.99.174303}, which is known to host a quantum Zeno crossover~\cite{PhysRevResearch.5.023204, PhysRevB.108.115127}: when $\gamma$ is small, the long-time relaxation rate of an initial state increases with $\gamma$; whereas when $\gamma$ is large, the long-time relaxation rate decreases with $\gamma$.  On the other hand, a finite $B$ is known to lead the Hamiltonian to exhibit Stark localization in both single-particle and many-body cases~\cite{RevModPhys.34.645, PhysRevLett.122.040606}, while it also destroys the solvability of the Liouvillian. The main object in the following is thus to introduce an efficient numerical method that uncovers the effect of Stark localization on Liouvillian dynamics.

Using dynamical correlators of the Liouvillian, we show the reduction of the relaxation rate with increasing $B$ for our model. In particular, we consider the relaxation of $\sigma^z_N$ above the steady state, which can be characterized by the following auto-correlator:
\beq \label{eq2}
C(t)=\text{tr}(\sigma^z_N(t)\sigma^z_N(0)\rho_s)=\text{tr}(\sigma^z_Ne^{t\mathcal{L}}[\sigma^z_N\rho_s])
\eeq
where $\rho_s$ is the density matrix of the steady state. Since our model only possesses Hermitian dissipators and conserves the parity operator $Q=\prod_{l=1}^N\sigma^z_l$, it has $2$ degenerate steady states $\rho_\pm=(\mathbf{I}\pm Q)/2^N$ where $\mathbf{I}$ denotes the identity matrix in the Pauli basis. Note that our method is applicable as long as $\rho_s$ is a low-entangled state after proper vectorization. We choose $\rho_s=(\rho_++\rho_-)/2=\mathbf{I}/2^N$ that fulfills this requirement in the following calculations~\footnote{The dynamical correlator for the steady state being other superposition of $\rho_\pm$ can be obtained via additionally computing $\tilde{C}(t)=\text{tr}(\sigma^z_Ne^{t\mathcal{L}}[\sigma^z_N Q])$, for which our method is still workable since $Q$ is also a low-entangled state after the same vectorization as Eq.~(\ref{eq3}).}.

To compute the correlator \eqref{eq2}, we vectorize a density matrix $\rho$ as
\beq \label{eq3}
\begin{aligned}
&\rho=\sum_{\sigma_1\cdots\sigma_N}\sum_{\tau_1\cdots\tau_N}\rho_{\sigma_1\cdots\sigma_N\tau_1\cdots\tau_N}|\sigma_1\cdots\sigma_N\rangle\langle \tau_1\cdots\tau_N|\\ &\to|\tilde{\rho}\rangle=\sum_{\sigma_1\cdots\sigma_N}\sum_{\tau_1\cdots\tau_N}\rho_{P[\sigma_1\cdots\sigma_N\tau_1\cdots\tau_N]}|\sigma_1\cdots\sigma_N\rangle| \tau_1\cdots\tau_N\rangle.
\end{aligned}
\eeq
where $\tau_1\cdots\tau_N$ are the degrees of freedom of the left vector in $\rho$, and $P[\tau_l]=\sigma_{2l}, P[\sigma_l]=\sigma_{2l-1}$ is a permutation of spin degrees of freedom on different sites. The advantage of implementing the permutation $P$ is that the transformed Liouvillian $\tilde{\mathcal{L}}$, which satisfies $\tilde{\mathcal{L}}|\tilde{\rho}\rangle=|\widetilde{\mathcal{L}[\rho]}\rangle$, will become a non-Hermitian spin model with only nearest and next-nearest neighbor interactions (see details in \appref{App_A}) [\figpanel{fig1}{b}]:
\beq \label{eq5}
\begin{aligned}
\tilde{\mathcal{L}}&=iJ_x\sum_{l=1}^{N/2}\left(\sigma^x_{4l-3}\sigma_{4l-1}^x-\sigma^x_{4l-2}\sigma_{4l}^x\right)\\&+ iJ_y\sum_{l=1}^{N/2-1}\left(\sigma^y_{4l-1}\sigma_{4l+1}^y-\sigma^y_{4l}\sigma_{4l+2}^y\right)\\&+iB\sum_{l=1}^N (l-1)\left(\sigma_{2l-1}^z-\sigma_{2l}^z\right)+\gamma\sum_{l=1}^{N}\sigma_{2l-1}^z\sigma_{2l}^z-\gamma N.
\end{aligned}
\eeq
In addition, the steady state transforms into $|\tilde{\rho}_s\rangle=\prod_{l=1}^N(|\uparrow_{2l-1}\uparrow_{2l}\rangle+|\downarrow_{2l-1}\downarrow_{2l}\rangle)/2$, for which only each two adjacent sites $2l-1$ and $2l$ are entangled. Thus, it is a low-entangled state as mentioned earlier. Lastly, after this vectorization, the inner product of two operators $\text{tr}(\rho_1^\dagger\rho_2)$ can be expressed as $\langle\tilde{\rho_1}|\tilde{\rho_2}\rangle$.

With this methodology, we can now compute \eqref{eq2} on the transformed basis
\beq \label{eq7}
C(t)=\langle\tilde{\mathbf{I}}|\sigma^z_{2N-1}e^{t\tilde{\mathcal{L}}}\sigma^z_{2N-1}|\tilde{\rho}_s\rangle,
\eeq
which can be obtained by integrating its frequency components
\beq \label{eq8}
C(t)=\int d^2\omega e^{\omega t}\mathcal{C}(\omega)\approx\sum_\omega(\Delta\omega)^2e^{\omega t}\mathcal{C}(\omega)
\eeq
with
\beq \label{eq9}
\mathcal{C}(\omega)=\langle\tilde{\mathbf{I}}|\sigma^z_{2N-1}\delta^2(\omega-\tilde{\mathcal{L}})\sigma^z_{2N-1}|\tilde{\rho}_s\rangle,
\eeq
where $\Delta\omega$ is the increment in $\omega$ when we discretize the integration into a summation. The Liouvilian is effectively a non-Hermitian operator, such that its dynamical correlator \eqref{eq9} can be computed with NHKPM~\cite{PhysRevLett.130.100401}, as detailed in \appref{App_B}. Moreover, the short-range nature of $\tilde{\mathcal{L}}$ as well as the low entanglement in $\sigma^z_{2N-1}|\tilde{\rho}_s\rangle$ favors the implementation of NHKPM using the matrix-product state (MPS) representation. 

\section{Results}
We first consider $N=4$, where the results are benchmarked with both exact diagonalization (ED) and the Runge-Kutta method in \appref{App_C}. For concreteness, we consider $J_x=0.75$, $J_y=0.5$ throughout the whole manuscript. With a fixed dissipation rate $\gamma=0.2$, $\mathcal{C}(\omega)$ in \eqref{eq9} is computed for $B=0,0.13$ and $0.25$ [\figpanels{fig1}{c}{e}], where we have presented the part of the spectrum featuring the peaks at the largest real frequency. We define $\Delta=|\text{max}(\Re(\omega_i))|$ where $\omega_i$ are the center of the peaks of $\mathcal{C}(\omega)$, such that $\Delta$ characterizes the relaxation rate of $C(t)$ at long times: $C(t)\sim e^{-\Delta t}$. For $B=0,0.13$ and $0.25$, we have $\Delta\approx0.52,0.53$ and $0.34$, respectively. Using \eqref{eq8}, we can now compute $C(t)$, as shown in \figpanel{fig1}{f} for short times and in \figpanel{fig1}{g} for long times and in a log scale. We see that the time evolution agrees with the spectrum, namely the long-time dynamics in all cases satisfy $C(t)\sim e^{-\Delta t}$.

\begin{figure}[t!]
\center
\includegraphics[width=\linewidth]{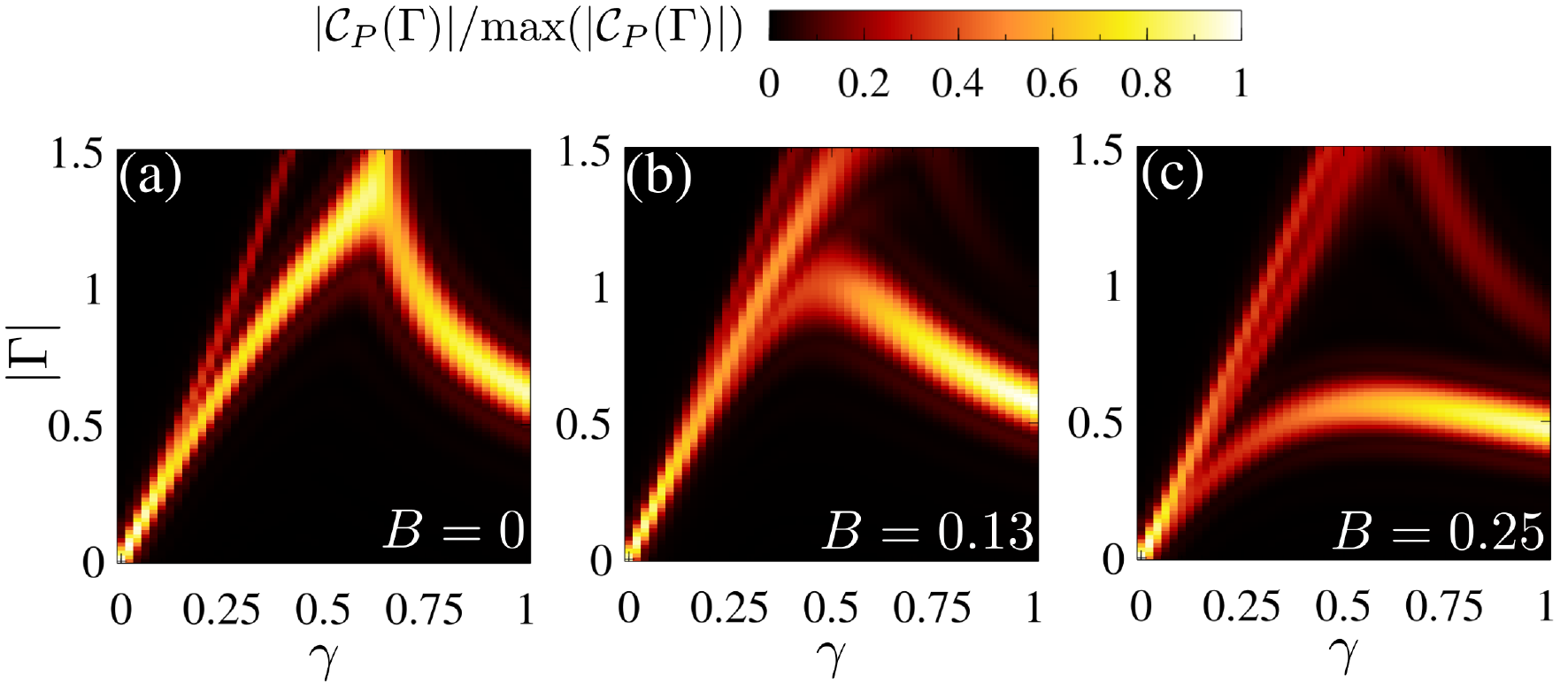}
\caption{The projected correlator of the Liouvillian $\mathcal{C}_P(\Gamma)$ in \eqref{eq10}, computed with $N=4$ and varying $\gamma$ for $B=0,0.13$ and $0.25$ in panels (a), (b) and (c), respectively; showing how the long-time relaxation rate $\Delta$ changes with $\gamma$ and $B$. Compared to $B=0$, $\Delta$ is decreased for $B=0.25$ when $\gamma$ is small due to Stark localization. However, this effect is hindered by the finite-size effect for $B=0.13$. }
\label{fig2}
\end{figure}

To see how the long-time relaxation rate $\Delta$ changes with $\gamma$, we compute the projected correlator
\beq \label{eq10}
\mathcal{C}_P(\Gamma)=\int d\Im \omega \mathcal{C}(\Gamma+i\Im \omega )\approx\sum_n \mathcal{C}(\Gamma+i\Im \omega_n) \Delta \omega,
\eeq
which projects the dynamical correlator $\mathcal{C}(\omega)$ to the real axis. The smallest $\Gamma$ at which $\mathcal{C}_P(\Gamma)$ exhibits a peak is thus equal to $\Delta$ by its definition. We note that the projected correlator is real as $\mathcal{C}(\Gamma+i\Im \omega)=\mathcal{C}^*(\Gamma-i\Im \omega)$, which is guaranteed by the universal property of Liouvillians $\mathcal{L}[\rho^\dagger]=(\mathcal{L}[\rho])^\dagger$ . The change of $\mathcal{C}_P(\Gamma)$ as a function of $\gamma$ for different $B$ are shown in \figref{fig2}. We observe that for $B=0$, $\Delta$ first increases linearly with $\gamma$, but then decreases after a critical point $\gamma_c\approx0.6$. This is known as the quantum Zeno crossover~\cite{PhysRevResearch.5.023204, PhysRevB.108.115127}. This crossover implies that the Liouvillian dynamics with a weak $\gamma$ and with a strong $\gamma$ are dominated by different physical factors. The properties of the Hamiltonian are crucial in the weak $\gamma$ regime, whereas the dissipative effects induced by dissipators take the main charge in the strong $\gamma$ regime. Consequently, the gradient magnetic field $B$ which can cause Stark localization in the Hamiltonian is expected to significantly influence the relaxation rate $\Delta$ within the weak $\gamma$ regime.

Increasing the magnetic field gradient $B$ is known to result in stronger localization of the eigenstates $\{|n\rangle\}$ of $H$~\cite{RevModPhys.34.645,PhysRevLett.122.040606}. In particular, when $B\gg J_{x,y}$, $|n\rangle$ will become localized enough to be also the eigenstates of all dissipators $L_l=\sigma_l^z$. In such a case, $\{|n\rangle\langle n|\}$ will span a decoherence-free space with zero relaxation rate. For a smaller $B$, the relaxation rate of this space is lifted due to the finite localization length of $|n\rangle$. As a result, the stronger the localization is, the smaller the relaxation rate becomes. This shows that Stark localization in the Hamiltonian can slow down the dynamics of the Liouvillian. A similar phenomenon was observed in a fermionic model~\cite{PhysRevLett.123.030602}. In our model, the reduction of the relaxation rate $\Delta$ in the weak $\gamma$ regime is observed for $B=0.25$ from both the auto-correlator $C(t)$ in \figpanel{fig1}{g} and the projected correlator $\mathcal{C}_P(\Gamma)$ in \figref{fig2}. However, for $B=0.13$, due to the finite-size effect, the localization length of the states is still larger than the system size $N=4$, and the reduction of the relaxation rate stemming from the localization is not seen.

\begin{figure}[t!]
\center
\includegraphics[width=\linewidth]{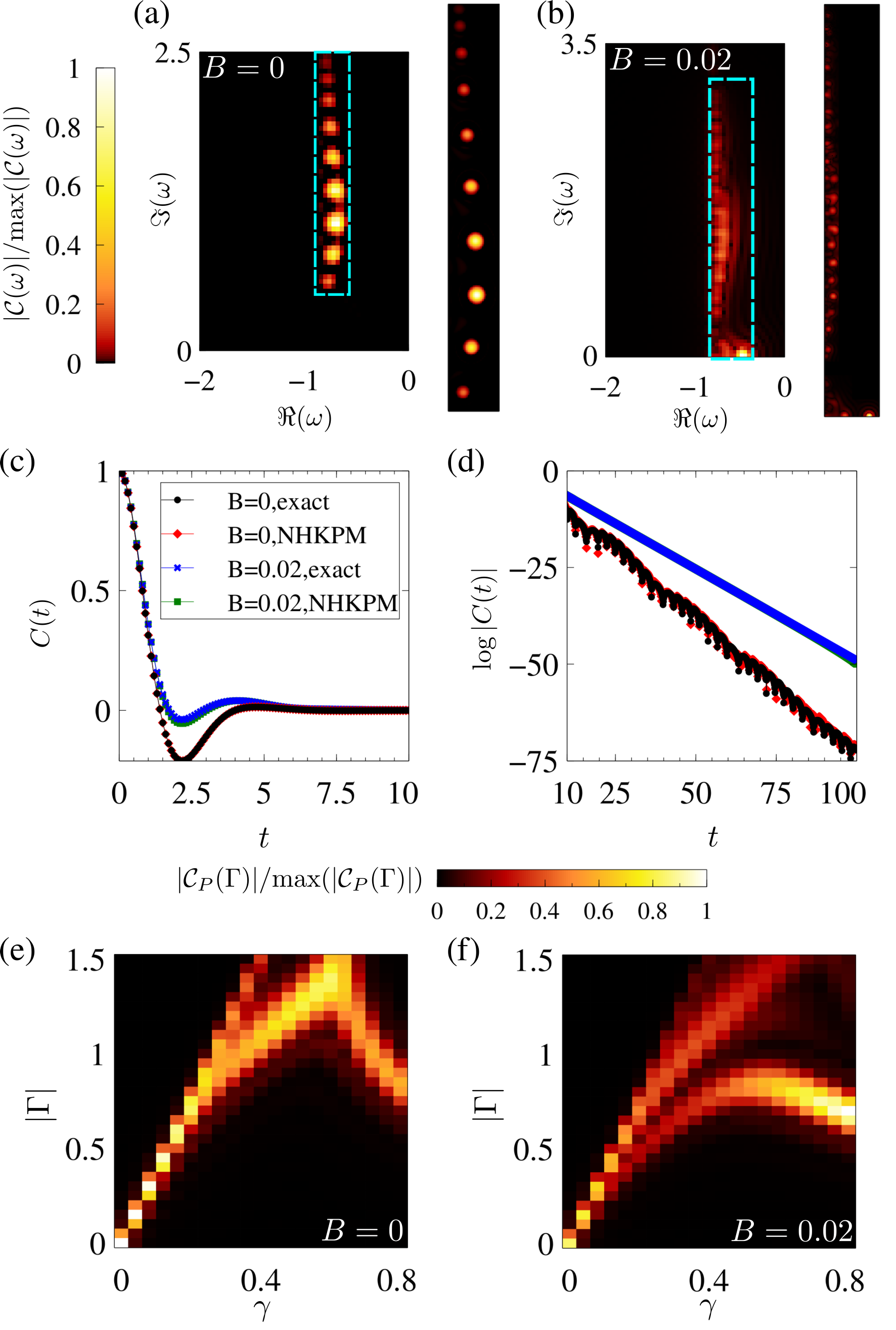}
\caption{The Liouvillian dynamics for $N=20$. (a,b) The dynamical correlator $\mathcal{C}(\omega)$ in \eqref{eq9} computed with $\gamma=0.2$ for $B=0$ and $B=0.02$, respectively. The insets show the $\mathcal{C}(\omega)$ computed with a higher resolution in the region specified with cyan rectangles. (c,d) The auto-correlator $C(t)$ in \eqref{eq7} for $B=0$ and $B=0.02$ computed exactly by diagonalizing the ``damping matrix" and with NHKPM. Panel (d) has the same labels as panel (c). Compared to $B=0$, the long-time relaxation rate $\Delta$ is decreased for $B=0.02$. (e,f) The projected correlator of the Liouvillian $\mathcal{C}_P(\Gamma)$ computed with varying $\gamma$ for $B=0$ and $B=0.02$, respectively. Compared to $B=0$, $\Delta$ is decreased for $B=0.02$ for small $\gamma$, showing that the finite-size effect has been eliminated compared to \figpanel{fig2}{b}.}
\label{fig3}
\end{figure}

To overcome finite-size effects, we now move on to consider a larger system size $N=20$. It is important that for our model, increasing the system size would not significantly shorten the weak $\gamma$ regime we are interested. This is different from the previously studied models with $U(1)$ symmetry~\cite{PhysRevLett.123.030602, PhysRevB.107.184303, PhysRevLett.132.070402}, for which the weak $\gamma$ regime shrinks to zero in the thermodynamic limit~\cite{PhysRevE.92.042143}. 
It is worth noting that, although our model with a finite $B$ is not solvable, its special structure still decouples the evolution of two-point correlators from other higher-order correlators. This property allows us to benchmark our results obtained from NHKPM by diagonalizing a ``damping matrix" of dimension $4N^2$ in \appref{App_D}.

We compute the autocorrelator $C(t)$ via combining the results of $\mathcal{C}(\omega)$ obtained from NHKPM with \eqref{eq8}. To reduce the computational cost, we first compute $\mathcal{C}(\omega)$ in a large area in the complex plane to identify the region where it shows peaks, and then compute it with a better frequency resolution in this specific region [insets in \figpanels{fig3}{a}{b}]. We see that NHKPM allows to faithfully extract $C(t)$, as the results agree with the exact results obtained by diagonalizing the ``damping matrix" [\figpanels{fig3}{c}{d}]. From \figpanel{fig3}{d}, it is obvious that even a rather small $B$ can lead to a slower long-time decay of $C(t)$ compared to $B=0$. To further demonstrate the decrease of $\Delta$ with increasing $B$ in the whole weak $\gamma$ regime, we show the projected correlator $\mathcal{C}_P(\Gamma)$ in \eqref{eq10} with varying $\gamma$ for both $B=0$ and $B=0.02$ [\figpanels{fig3}{e}{f}]. In contrast to the case of $N=4$ and $B=0.13$, in the case of $N=20$ and $B=0.02$, the decrease of $\Delta$ compared to $B=0$ is observed, showing that in this case finite-size effects are eliminated.

\begin{figure}[t!]
\center
\includegraphics[width=\linewidth]{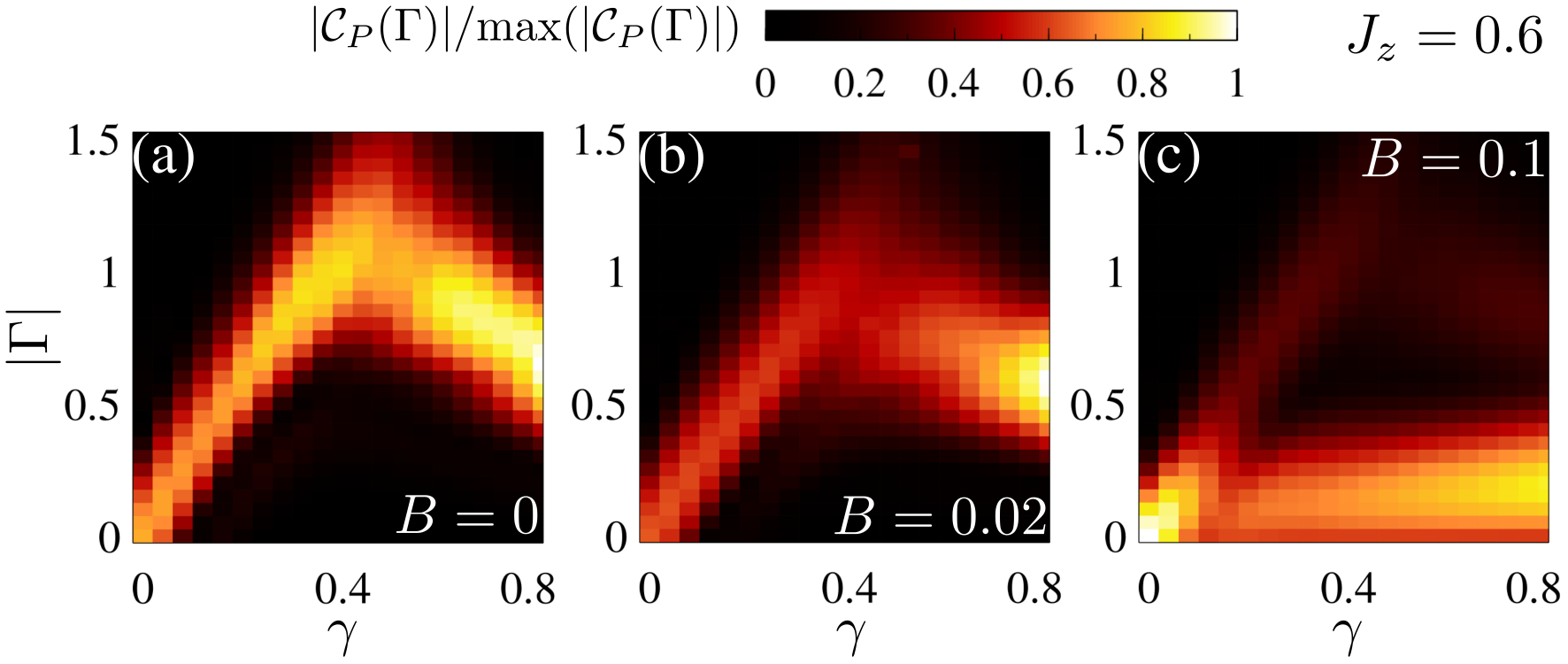}
\caption{The projected correlator of the Liouvillian $\mathcal{C}_P(\Gamma)$ in \eqref{eq10}, computed with $N=20$ and $J_z=0.6$; $B=0$ in (a), $B=0.02$ in (b) and $B=0.1$ in (c). (a) The presence of a finite $J_z$ reduces the quantum Zeno crossover point to around $\gamma_c\approx0.4$. (b) The reduction of $\Delta$ in the presence of $B=0.02$ is less significant compared to \figpanel{fig3}{f} due to a finite threshold in $B$ to cause Stark localization. (c) For a larger $B=0.1$, the reduction of $\Delta$ becomes visible.}
\label{fig4}
\end{figure}

Finally, we include an interaction term to the Hamiltonian of the form $\sum_{l=1}^{N-1}J_z\sigma^z_l\sigma^z_{l+1}$. With this interaction term, it is no longer possible to solve the spectrum and dynamics with exact methods, while NHKPM can still be applied. We are particularly interested in how this interaction influences the quantum Zeno crossover and the reduction of the relaxation rate due to Stark localization. For concreteness, we take $J_z=0.6$ in our analysis. We observe that when $B=0$, the quantum Zeno crossover point $\gamma_c$ is reduced compared to the case $J_z=0$ [\figpanel{fig4}{a}]. This reduction of $\gamma_c$ can be understood from the fact that when $J_z\to\infty$, the model will have $U(1)$ symmetry, resulting in $\gamma_c=0$ in the thermodynamic limit~\cite{PhysRevLett.123.030602, PhysRevB.107.184303, PhysRevLett.132.070402}; a finite $J_z$ does not fully restore this symmetry, but can still reduce $\gamma_c$. 
Stark localization in non-interacting systems takes place with an infinitesimal gradient potential.
In contrast, the introduction of the interaction $\sum_{l=1}^{N-1}J_z\sigma^z_l\sigma^z_{l+1}$ leads to Stark many-body localization only when the gradient $B$ is above a finite threshold~\cite{doi:10.1073/pnas.1819316116}. This is consistent with our numerical findings, as the reduction of the relaxation with a small $B=0.02$ is not significant, as shown in [\figpanel{fig4}{b}].
However, such relaxation becomes visible for a larger $B=0.1$ [\figpanel{fig4}{c}].

\section{Conclusion}
We have demonstrated a method to solve many-body Liouvillian dynamics based on NHKPM and tensor-network techniques. This method allows to compute the dynamical correlator of a many-body Liouvillian, providing access to the Liouvillian spectrum and Liouvillian dynamics. Focusing on the dephasing quantum compass model with a gradient magnetic field, we have demonstrated the faithfulness of our method by comparing it with exact results. This enabled the characterization of the quantum Zeno crossover and the reduction of relaxation rate due to Stark localization in our model. We further demonstrated the capabilities of our method in regimes
where previous exact methods cannot be applied, in particular
by exploring how nearest-neighbor
interactions influence the Liouvillian dynamics of our model.
Compared to existing tensor-network methods for open quantum dynamics such as TEBD~\cite{PhysRevLett.91.147902, PhysRevLett.93.207205, PhysRevB.102.064304} and TDVP~\cite{PhysRevLett.107.070601, PhysRevB.94.165116}, our method focuses on the computation of the dynamical correlator $\mathcal{C}(\omega)$ instead of the auto-correlator $C(t)$. The auto-correlator $C(t)$ can be determined by the dynamical correlator $\mathcal{C}(\omega)$, while the converse is not true. In this sense, our method can provide more information to explore the underlying mechanisms of the open quantum dynamics. Specifically, the long-time relaxation rate can be directly extracted from $\mathcal{C}(\omega)$. Additionally, our method is particularly favorable for computing long-time dynamics as the computational cost of $C(t)$ from $\mathcal{C}(\omega)$ does not increase with $t$.
Our method can be applied to a variety of open quantum many-body systems with short-range interactions~\cite{PhysRevB.102.064304, PhysRevLett.132.120401, PhysRevB.105.205125, PhysRevLett.107.137201, PhysRevLett.112.030603}. The dynamical correlators computed with our method would allow to explore interesting properties such as the Liouvillian gap and novel non-Hermitian topology of these systems. Finally, beyond its fundamental interest, our methodology
would allow rationalizing dynamical correlators at complex frequencies measured experimentally~\cite{PhysRevLett.129.093001,PhysRevLett.130.163001}.

\textbf{Acknowledgements}
We thank Sebastian Diehl and Hosho Katsura for fruitful discussions. We acknowledge the computational resources provided by the Aalto Science-IT project. Guangze Chen is supported by European Union's Horizon 2023 research and innovation programme under the Marie Skłodowska-Curie grant agreement No. 101146565, J. L. Lado acknowledges financial support from the Academy of Finland Projects No.331342 and No. 358088 and the Jane and Aatos Erkko Foundation, and Fei Song is supported by NSFC under Grant No.~12404189 and the Postdoctoral Fellowship Program of CPSF under Grant No. GZB20240732. The ITensor library~\cite{itensor} has been used in the numerical calculations. 


\appendix

\section{Vectorization of a density matrix} \label{App_A}
In this section, we explain the choice of basis for the vectorization of a density matrix. The most intuitive way of vectorizing $\rho$ for our model $\mathcal{L}$ is
\beq \label{eqS1}
\begin{aligned}
&\rho=\sum_{\sigma_1\cdots\sigma_N}\sum_{\tau_1\cdots\tau_N}\rho_{\sigma_1\cdots\sigma_N\tau_1\cdots\tau_N}|\sigma_1\cdots\sigma_N\rangle\langle \tau_1\cdots\tau_N|\\ &\to|\tilde{\rho}\rangle=\sum_{\sigma_1\cdots\sigma_N}\sum_{\tau_1\cdots\tau_N}\rho_{\sigma_1\cdots\sigma_N\tau_1\cdots\tau_N}|\sigma_1\cdots\sigma_N\rangle| \tau_1\cdots\tau_N\rangle.
\end{aligned}
\eeq
Under this basis choice, the transformed Liouvillian is
\beq \label{eqS2}
\begin{aligned}
\tilde{\mathcal{L}}&=-iH\otimes\mathbf{I}+i\mathbf{I}\otimes H^T\\&+\sum_l\left(L_l\otimes L_l^*-\frac{1}{2}(L^\dag_l L_l\otimes\mathbf{I}+\mathbf{I}\otimes L_l^T L_l^*)\right)\\
&=iJ_x\sum_{l=1}^{N/2}\left(\sigma^x_{2l-1}\sigma_{2l}^x-\sigma^x_{2l-1+N}\sigma_{2l+N}^x\right)\\&+iJ_y\sum_{l=1}^{N/2-1}\left(\sigma^y_{2l}\sigma_{2l+1}^y-\sigma^y_{2l+N}\sigma_{2l+1+N}^y\right)\\&+iB\sum_{l=1}^N (l-1)\left(\sigma_{l}^z-\sigma_{l+N}^z\right)\\&+\gamma\sum_{l=1}^{N}\sigma_{l}^z\sigma_{l+N}^z-\gamma N.
\end{aligned}
\eeq
where $H$ and $L_l$ are given in Eq.(1) in the main text. We see that in \eqref{eqS2} long-range interactions $\sigma_{l}^z\sigma_{l+N}^z$ exist, in particular when $N$ is large. This is not favorable for calculations with tensor-networks. In addition, under this basis, the steady state $\rho_s=\mathbf{I}/2^N$ transforms into:
\beq
|\tilde{\rho}_s\rangle=\frac{1}{2^N}\sum_{\sigma_1\cdots\sigma_{2N}}\delta_{\sigma_1\sigma_{N+1}}\delta_{\sigma_2\sigma_{N+2}}\cdots\delta_{\sigma_{N}\sigma_{2N}}|\sigma_1\cdots\sigma_{2N}\rangle
\eeq
which lacks a simple matrix-product-state (MPS) representation. For both reasons, we choose the basis in \eqref{eq3} for vectorization where spin indices have been permuted.


\section{Dynamical correlators with NHKPM} \label{App_B}

From our previous work~\cite{PhysRevLett.130.100401}, We derived that an arbitrary dynamical correlator
\beq\label{eqSS1}
f(\omega)=\langle \psi_\text{L}|\delta^2(\omega-\tilde{\mathcal{L}})|\psi_\text{R}\rangle
\eeq
is equal to 
\beq
\begin{aligned}
f(\omega)&=\partial_{\omega^*}\langle \psi_\text{L}|(\omega-\tilde{\mathcal{L}})^{-1}|\psi_\text{R}\rangle\\
&=\sum_n\partial_{\omega^*}(\omega-\omega_n)^{-1}\langle \psi_\text{L}|\psi_\text{R,n}\rangle\langle \psi_\text{L,n}|\psi_\text{R}\rangle
\end{aligned}
\eeq
where $\omega_n$ and $|\psi_\text{R(L),n}\rangle$ are the $n$th eigenvalue and right (left) eigenvector of $\tilde{\mathcal{L}}$. Thus, \eqref{eqSS1} can be computed with
\beq
f(\omega)=\frac{1}{\pi}\partial_{\omega^*}G(E=0)
\eeq
where
\beq
G(E)=\langle\text{L}|(E-{\mathcal{H}})^{-1}|\text{R}\rangle
\eeq
is an entry of the Green's function of the Hermitrized Hamiltonian $\mathcal{H}$:
\beq \label{eq_B5}
\mathcal{H}=\left(\begin{array}{cc}&\omega-\tilde{\mathcal{L}}\\\omega^*-\tilde{\mathcal{L}}^\dag&\end{array}\right)
\eeq
with
\beq
|\text{L}\rangle=\left(\begin{array}{c}0\\|\psi_\text{L}\rangle\end{array}\right),|\text{R}\rangle=\left(\begin{array}{c}|\psi_\text{R}\rangle\\0\end{array}\right).
\eeq
Since $G(E)$ is a function of a single variable, we can apply the kernel polynomial method (KPM) to compute $G(E)$, giving
\beq
G_\text{KPM}(E=0)=\sum_n\langle\text{L}|\varphi_n\rangle\left(\mathcal{E}_n^{-1}\right)_\text{KPM}\langle\varphi_n|\text{R}\rangle
\eeq
where $\mathcal{E}_n$ and $|\varphi_n\rangle$ are the $n$th eigenvalue and eigenvector of $\mathcal{H}$. The approximated function $\left(\mathcal{E}_n^{-1}\right)_\text{KPM}$ depends on the kernel function and the number of polynomials in the Chebyshev expansion~\cite{RevModPhys.78.275}. In particular, for calculations in the main text, we have chosen the Jackson kernel, resulting in:
\beq \label{eqS28}
\left(\frac{1}{\mathcal{E}_n}\right)_\text{J}\approx\frac{2}{\sqrt{2\sigma^2}}F\left(\frac{\mathcal{E}_n}{\sqrt{2\sigma^2}}\right)
\eeq
where $F(x)=\exp(-x^2)\int_0^x\exp(t^2)dt$ is the Dawson function~\cite{https://doi.org/10.1112/plms/s1-29.1.519} and $\sigma=\pi/N$ where $N$ is the number of polynomials in the Chebyshev expansion. \eqref{eqS28} provides a good approximation for $1/\mathcal{E}_n$ for $\mathcal{E}_n\gtrsim2\sigma$:
\beq \label{eqS14}
\frac{2}{\sqrt{2\sigma^2}}F\left(\frac{\mathcal{E}_n}{\sqrt{2\sigma^2}}\right)=\frac{1}{\mathcal{E}_n}+O\left(\frac{\sigma^2}{\mathcal{E}_n^3}\right) \quad (\mathcal{E}_n\gtrsim2\sigma)
\eeq

From \eqref{eq_B5} we see that $\mathcal{E}_n=\pm|\omega-\omega_n|$, where $\omega_n$ are the eigenvalues of $\tilde{\mathcal{L}}$. Thus, the KPM approximates $|\omega-\omega_n|^{-1}$ with $(|\omega-\omega_n|^{-1})_\text{J}$:
\beq \label{eq15}
\begin{aligned}
&f_\text{KPM}(\omega)\\=&\sum_n\partial_{\omega^*}\mleft[\mleft(\frac{1}{|\omega-\omega_n|}\mright)_\text{J}e^{-i\arg(\omega-\omega_n)}\mright]\langle \psi_\text{L}|\psi_\text{R,n}\rangle\langle \psi_\text{L,n}|\psi_\text{R}\rangle\\
=&\sum_n\partial_{\omega^*}\mleft[\mleft(\frac{1}{|\omega-\omega_n|}\mright)_\text{J}\frac{|\omega-\omega_n|}{\omega-\omega_n}\mright]\langle \psi_\text{L}|\psi_\text{R,n}\rangle\langle \psi_\text{L,n}|\psi_\text{R}\rangle.
\end{aligned}
\eeq

We now demonstrate that $f_\text{KPM}(\omega)$ provides a good approximation to $f(\omega)$ upon integration. Let
\beq
g(\omega-\omega_n)=\mleft(\frac{1}{|\omega-\omega_n|}\mright)_\text{J}\frac{|\omega-\omega_n|}{\omega-\omega_n}
\eeq
and note that for a region $\Sigma$ in the complex plane:
\beq \label{eqS12}
\begin{aligned}
&\int_\Sigma d^2\omega \partial_{\omega^*}g(\omega-\omega_n)\\
=&\frac{1}{2}\int_\Sigma d^2\omega \left(\partial_xg(\omega-\omega_n)-\partial_y\left(ig(\omega-\omega_n)\right)\right)\\
=&\frac{1}{2}\int_{d\Sigma}\left(g(\omega-\omega_n),ig(\omega-\omega_n)\right)\cdot d\mathbf{l}
\end{aligned}
\eeq
where we have used the Stokes theorem. From \eqref{eqS14} we see that as long as $|\omega-\omega_n|>2\sigma$ when $\omega\in d\Sigma$, $g(\omega-\omega_n)\approx(\omega-\omega_n)^{-1}(1+O(\sigma^2|\omega-\omega_n|^{-2}))$ will also hold for $\omega\in d\Sigma$, and \eqref{eqS12} reduces to
\beq
\begin{aligned}
&\int_\Sigma d^2\omega \partial_{\omega^*}g(\omega-\omega_n)\\
\approx&\frac{1}{2}\int_{d\Sigma}\left((\omega-\omega_n)^{-1},i(\omega-\omega_n)^{-1}\right)\cdot d\mathbf{l}\\
=&\int_\Sigma d^2\omega \partial_{\omega^*}(\omega-\omega_n)^{-1}.
\end{aligned}
\eeq
Thus, when a complex plane $\Sigma$ satisfy: $|\omega-\omega_n|>2\sigma$, $\forall \omega\in d\Sigma, \forall n$, we will have
\beq \label{eqS19}
\int_\Sigma d^2\omega f_\text{KPM}(\omega)\approx\int_\Sigma d^2\omega f(\omega).
\eeq
The error decreases as $\Sigma$ becomes larger, and in particular, if $\Sigma$ is the whole complex plane then both sides of \eqref{eqS19} are equal. Note that when $\sigma$ is small enough, which can be achieved with a larger number of polynomials $N$, \eqref{eqS19} will hold in a small region containing only one specific eigenvalue $\omega_i$. This means that the spectral weight of this eigenvalue computed with NHKPM is equal to its exact value.

We have thus shown that, NHKPM not only allows to predict the eigenvalues of $\tilde{\mathcal{L}}$, as demonstrated in our previous work~\cite{PhysRevLett.130.100401}, but it also correctly predicts the spectral weight of the eigenvalues. This essentially enables the calculation of the auto-correlator from dynamical correlators with \eqref{eq8} in the main text.

\begin{figure}[t!]
\center
\includegraphics[width=\linewidth]{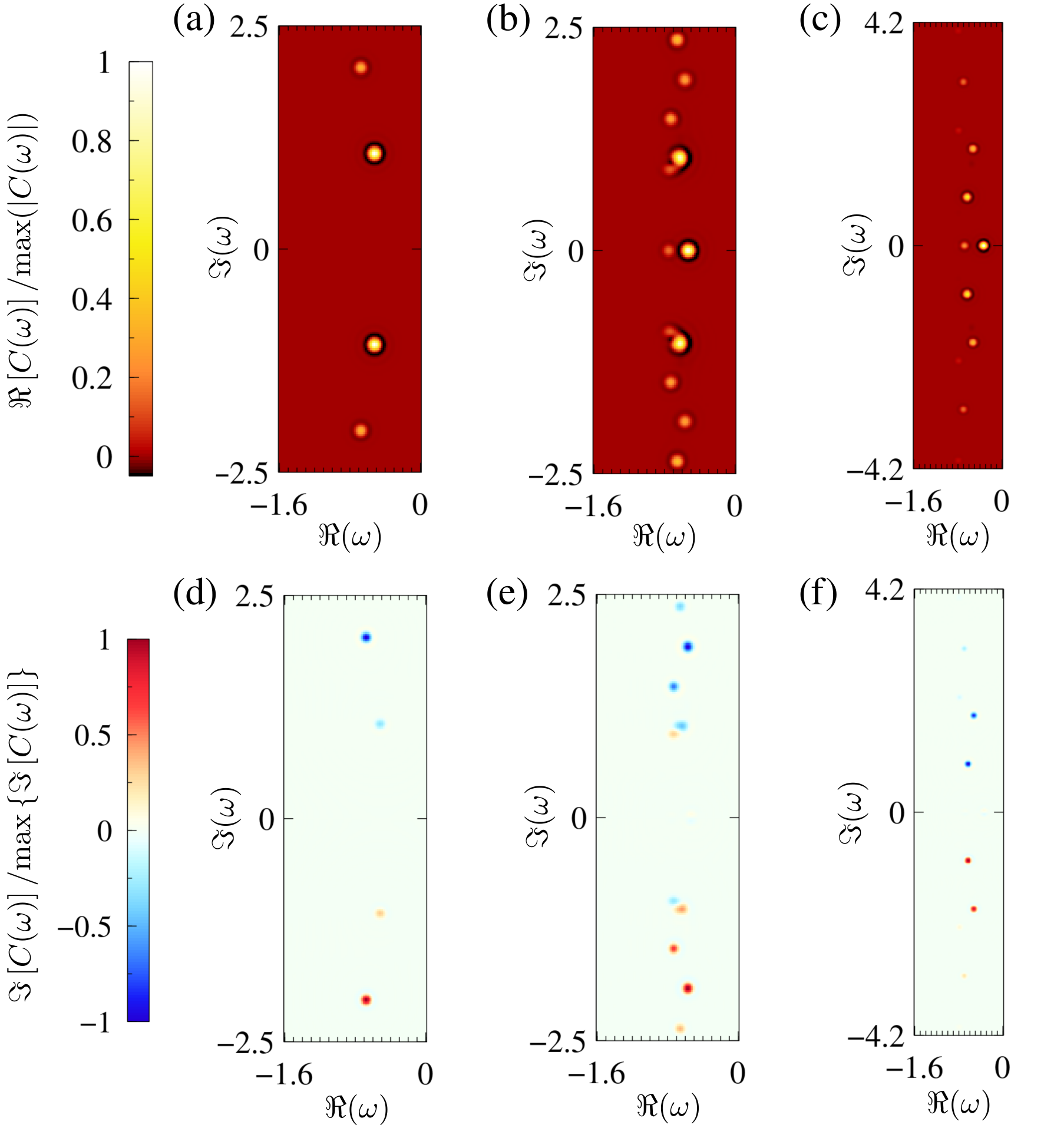}
\caption{Real and imaginary parts of the dynamical correlator $\mathcal{C}(\omega)$ in Fig.1(c)-(e) in the main text. Panels (a) and (d) are computed with $B=0$, panels (b) and (e) are computed with $B=0.13$, and Panels (c) and (f) are computed with $B=0.25$. We see that the computed $\mathcal{C}(\omega)$ satisfy $\mathcal{C}(\omega^*)=\mathcal{C}^*(\omega)$.
}
\label{figS1}
\end{figure}

We now move on to show that NHKPM maintains an important property of the dynamical correlator $\mathcal{C}(\omega)$ that ensures the auto-correlator $C(t)$ to be real. The auto-correlator is computed as:
\beq \label{eq_S18}
C(t)=\int d^2\omega e^{\omega t}\mathcal{C}(\omega),
\eeq
where
\beq
\mathcal{C}(\omega)=\langle\tilde{\mathbf{I}}|\sigma^z_{2N-1}\delta^2(\omega-\tilde{\mathcal{L}})\sigma^z_{2N-1}|\tilde{I}\rangle/2^N
\eeq
We note that, the eigen-operators of $\mathcal{L}$ are either Hermitian or in pairs of Hermitian conjugation conjugation~\cite{PhysRevA.98.042118}:
\begin{equation}
\mathcal{L}[\rho_n^\dag]=\omega_n^*\rho_n^\dag \text{ if } \mathcal{L}[\rho_n]=\omega_n\rho_n,
\end{equation}
or in the transformed form
\beq \label{eqS23}
\tilde{\mathcal{L}}|\widetilde{\rho_n^\dag}\rangle=\omega_n^*|\widetilde{\rho_n^\dag}\rangle \text{ if } \tilde{\mathcal{L}}|\tilde{\rho}_n\rangle=\omega_n|\tilde{\rho}_n\rangle.
\eeq
As a consequence, if $\rho$ is Hermitian, we have:
\beq \label{eqS24}
\begin{aligned}
&\langle \tilde{\rho}|\delta^2(\omega-\tilde{\mathcal{L}})|\tilde{\rho}\rangle\\
=&\sum_n\delta^2(\omega-\omega_n^*)\langle \tilde{\rho}|\widetilde{\rho_n^\dag}\rangle\langle\widetilde{\rho_{\text{L},n}^\dag}|\tilde{\rho}\rangle\\
=&\sum_n\delta^2(\omega-\omega_n^*)\langle \widetilde{\rho^\dag}|\widetilde{\rho_n^\dag}\rangle\langle\widetilde{\rho_{\text{L},n}^\dag}|\widetilde{\rho^\dag}\rangle\\
=&\left(\sum_n\delta^2(\omega-\omega_n^*)\langle \tilde{\rho}|\tilde{\rho}_n\rangle\langle\tilde{\rho}_{\text{L},n}|\tilde{\rho}\rangle\right)^*\\
=&\left(\langle \tilde{\rho}|\delta^2(\omega-\tilde{\mathcal{L}})|\tilde{\rho}\rangle\right)^*
\end{aligned}
\eeq
where $\langle\tilde{\rho}_{\text{L},n}|$ is the left eigenvector of $\tilde{\mathcal{L}}$ corresponding to eigenvalue $\omega_n$. We have used \eqref{eqS23} in deriving the second line of \eqref{eqS24}. Since $\rho=\sigma_N^z\mathbf{I}$ is Hermitian, we have
\beq
\mathcal{C}(\omega^*)=\mathcal{C}^*(\omega),
\eeq
an important property that ensures $C(t)$ is real. We see that the results obtained with NHKPM maintain this property (\figref{figS1}).

\section{Dynamics with ED and Runge-Kutta method for $N=4$} \label{App_C}
\begin{figure}[t!]
\center
\includegraphics[width=\linewidth]{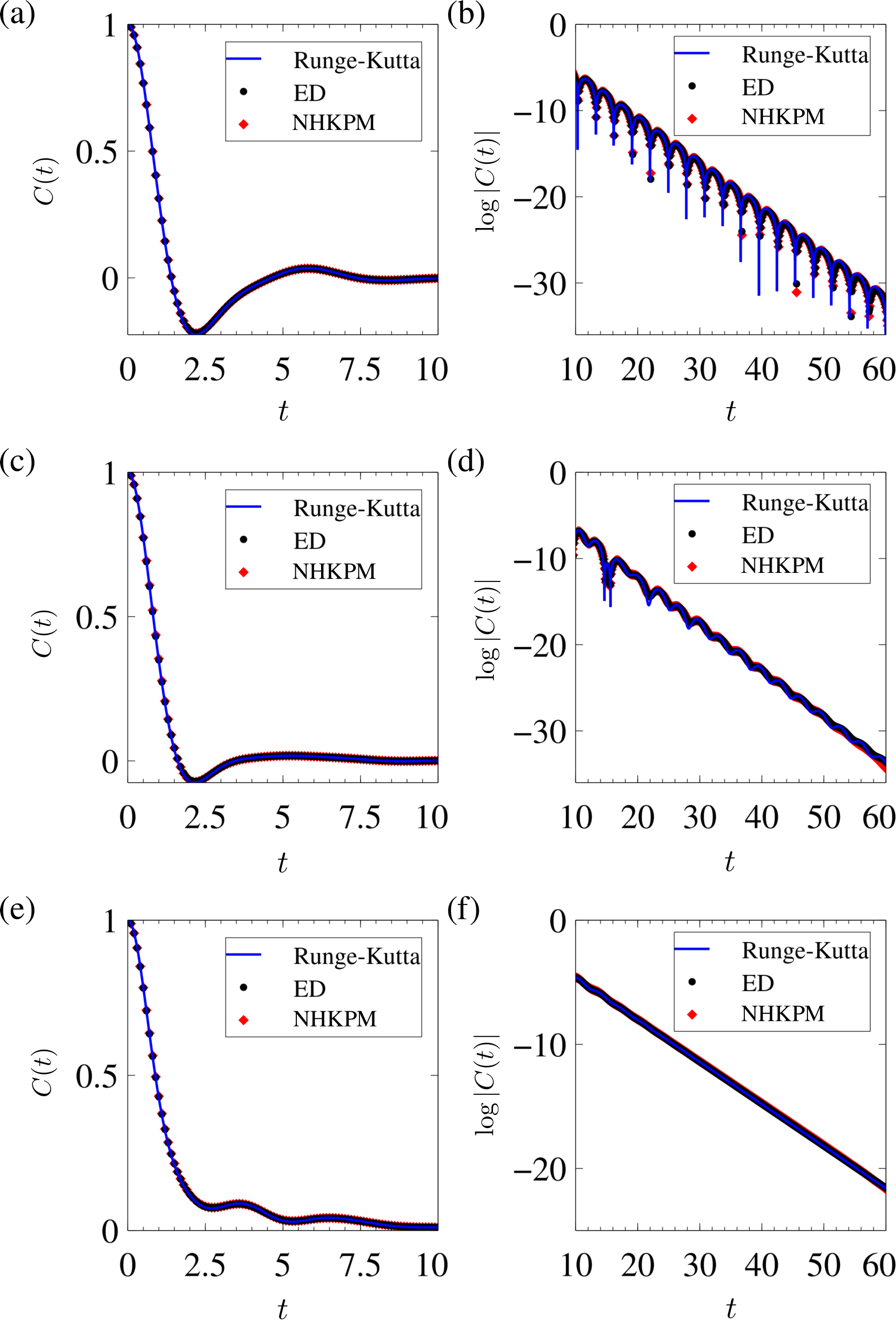}
\caption{Benchmark of $C(t)$ in Fig.1(f)-(g) in the main text. Panels (a) and (b) are computed with $B=0$, panels (c) and (d) are computed with $B=0.13$, and Panels (e) and (f) are computed with $B=0.25$. The results with both ED and Runge-Kutta methods agree with the results obtained with NHKPM.}
\label{figS2}
\end{figure}

We present the numerical details for computing the auto-correlator $C(t)$ with exact diagonalization (ED) and the Runge-Kutta method. We note that
\beq \label{eqS26}
C(t)=\text{tr}(\sigma^z_N(t)\sigma^z_N(0)\rho_s)=\text{tr}(\sigma^z_Ne^{t\mathcal{L}}[\sigma^z_N\rho_s])
\eeq
when the Liouvillian is treated as a superoperator, and
\beq \label{eqS27}
C(t)=\langle\tilde{\mathbf{I}}|\sigma^z_{2N-1}e^{t\tilde{\mathcal{L}}}\sigma^z_{2N-1}|\tilde{\rho}_s\rangle,
\eeq
in the vectorized form. 

To compute $C(t)$ with ED, we use \eqref{eqS27}, and diagonalize $\tilde{\mathcal{L}}$:
\beq
\tilde{\mathcal{L}}=\sum_n\omega_n|\tilde{\rho}_n\rangle\langle\tilde{\rho}_{n,L}|
\eeq
such that $C(t)$ can be computed with
\beq
C(t)=\sum_n\langle\tilde{\mathbf{I}}|\sigma^z_{2N-1}|\tilde{\rho}_n\rangle e^{\omega_n t}\langle\tilde{\rho}_{n,L}|\sigma^z_{2N-1}|\tilde{\rho}_s\rangle.
\eeq

To compute $C(t)$ with the Runge-Kutta method, we can directly use \eqref{eqS26} without the need for vectorization. Let
\beq
y(t)=e^{t\mathcal{L}}[\sigma^z_N\rho_s],
\eeq
we have
\beq
\frac{dy}{dt}=\mathcal{L}[y(t)]\equiv f(y)
\eeq
with $y(t=0)=\sigma^z_N\rho_s$. This allows us to use the Runge-Kutta method~\cite{Runge1895} to iteratively compute $y(t)$ with a time step of $h$:
\beq
\begin{aligned}
y_{n+1}&=y_n+\frac{h}{6}(k_1+2k_2+2k_3+k_4)\\
t_{n+1}&=t_n+h
\end{aligned}
\eeq
where
\beq
\begin{aligned}
k_1&=f(y_n)\\
k_2&=f(y_n+h\frac{k_1}{2})\\
k_3&=f(y_n+h\frac{k_2}{2})\\
k_4&=f(y_n+hk_3)\\
\end{aligned}
\eeq
with $y_0=\sigma^z_N\rho_s$ and $t_0=0$.

The computations of $C(t)$ in \figpanels{fig1}{f}{g} in the main text with both ED and Runge-Kutta method are shown in \figref{figS2}, where both results verify the correctness of NHKPM in computing $C(t)$.

\section{Dynamics obtained using the closed hierarchy of correlations for $N=20$} \label{App_D}
\begin{figure}[t!]
\center
\includegraphics[width=\linewidth]{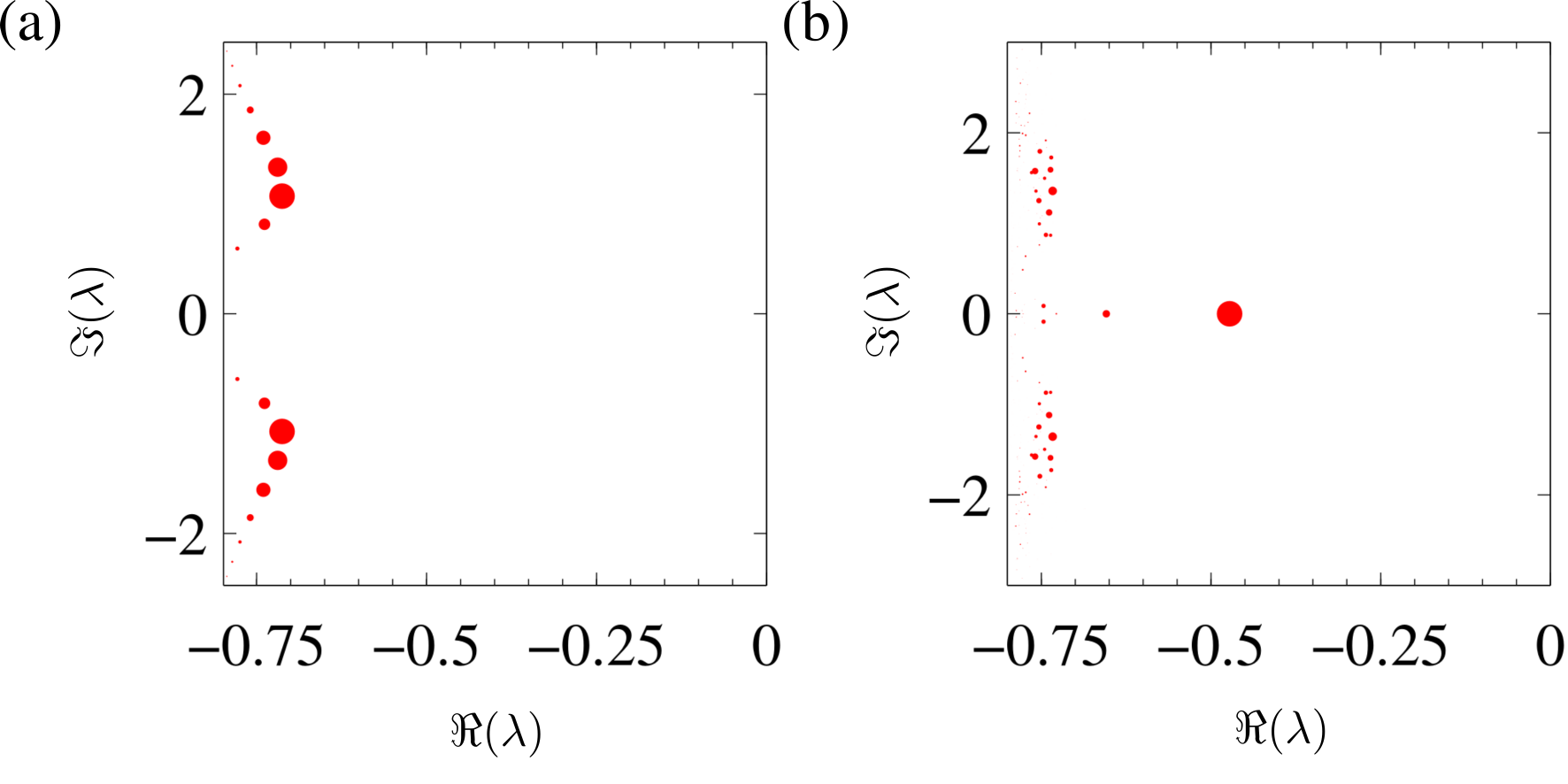}
\caption{Spectrum of the damping matrix $X$ in \eqref{eqS49} with $N=20$ and $B=0$ in panel (a), and $B=0.02$ in panel (b). The spectrum is highlighted with red color, and the size of the dots is proportional to $\langle nL|\tilde{o}(0)\rangle$, the overlap between the corresponding eigenstate of the eigenvalue and the state $|\tilde{o}(0)\rangle$.}
\label{figS3}
\end{figure}

To compute $C(t)$ in the case of $N=20$, both ED and the Runge-Kutta method fail due to the exponentially large Hilbert space. However, in our model, the Liouvillian preserves two-point correlators, allowing us to compute the dynamics by diagonalizing a "damping matrix"~\cite{PhysRevB.109.L140302, barthel2022solving}. We present the numerical details below.

It is simpler to deal with the quantum compass model under the Majorana basis~\cite{PhysRevB.99.174303}. This is done by performing the Jordan-Wigner transformation~\cite{Jordan1928}
\beq
\begin{aligned}
\sigma_l^+&=\prod_{j<l}e^{i\pi n_j} c_l^\dag\\
\sigma_l^-&=c_l\prod_{j<l}e^{-i\pi n_j} \\
\sigma_l^z&=2c_l^\dag c_l-1
\end{aligned}
\eeq
to transform the spin model into a fermion model, followed by a transformation into the following Majorana basis:
\beq
\begin{aligned}
&\gamma_l^-=i(c_i-c_i^\dag),\quad \gamma_l^+=(c_i+c_i^\dag)\quad (l\quad \text{odd})\\
&\gamma_l^-=(c_i+c_i^\dag),\quad \gamma_l^+=i(c_i-c_i^\dag)\quad (l\quad \text{even})
\end{aligned}
\eeq
where it can be verified $\{\gamma_l^\alpha,\gamma_j^\beta\}=2\delta_{\alpha\beta}\delta_{lj}$. An important property of the Majorana operators is:
\beq
[\gamma_i\gamma_j,\gamma_l\gamma_m]=2\gamma_i\gamma_m\delta_{jl}-2\gamma_i\gamma_l\delta_{jm}+2\gamma_m\gamma_j\delta_{il}-2\gamma_l\gamma_j\delta_{im},
\eeq
which ensures the commutator of two quadratic operators is still quadratic. A Hermitian quadratic operator $A$ under the Majorana basis is:
\beq
A=\frac{i}{4}\vec{\gamma}^Ta\vec{\gamma}
\eeq
with $a=-a^T$, where \beq
\vec{\gamma}=\left(\begin{array}{c}\gamma_1^-\\ \vdots\\ \gamma_N^-\\ \gamma_1^+\\ \vdots\\ \gamma_N^+\end{array}\right).
\eeq

In particular, in our model, both the Hamiltonian $H$ and the Lindblad dissipators are Hermitian, which ensures the preservation of two-point correlators. Furthermore, the density matrix $O=\sigma_N^z\rho_\text{ss}$ is also Hermitian. Thus, the time evolution can be computed in the single-particle subspace of Majorana fermions. In the Majorana basis, the Hamiltonian $H$ can be expressed as
\beq
\begin{aligned}
H=\frac{i}{4}\vec{\gamma}^Th\vec{\gamma}
\end{aligned}
\eeq
where
\beq
h=\left(\begin{array}{cc}T & M\\ -M & \end{array}\right)
\eeq
with
\beq
T=\left(\begin{array}{cccc}0 & 2J_x&0&0\\ -2J_x&0&-2J_y&0\\0&2J_y&0&2J_x\\ 0& 0&\ddots&\ddots\end{array}\right)
\eeq
and
\beq
M=\left(\begin{array}{cccc}0\\&2B\\&&4B\\&&&\ddots\end{array}\right).
\eeq
Similarly, the Lindblad dissipator $L_l=\sqrt{\gamma}\sigma_l^z$ becomes
\beq
L_l=\frac{i}{4}\vec{\gamma}^Tl_l\vec{\gamma}
\eeq
where
\beq
l_l=\left(\begin{array}{cc}0 & \tilde{l}_l\\ -\tilde{l}_l & 0\end{array}\right)
\eeq
with
\beq
\left(\tilde{l}_l\right)_{ij}=2\sqrt{\gamma}\delta_{il}\delta_{jl}.
\eeq
Finally, the density matrix $O$ can also be expanded
\beq
O=\frac{i}{4}\vec{\gamma}^To\vec{\gamma}.
\eeq

When the dissipators $L_l$ are Hermitian, the quantum master equation becomes
\beq
\frac{d}{dt}O=-i[H,O]-\frac{1}{2}\sum\left[L_l,[L_l,O]\right].
\eeq
Under the Majorana basis, this is
\beq \label{eqS48}
\begin{aligned}
\frac{d}{dt}o&=[h,o]-\frac{1}{2}\sum\left[l_l,[l_l,o]\right]\\
&=ho-oh-\sum_ll_lol_l-4\gamma o.
\end{aligned}
\eeq
In the vectorized form, \eqref{eqS48} becomes
\beq \label{eqS49}
\begin{aligned}
\frac{d}{dt}|\tilde{o}\rangle&=\left[\left(h\otimes\mathbf{1}-\mathbf{1}\otimes h^T-\sum_ll_l\otimes l_l^T\right)-4\gamma\mathbf{1}\otimes\mathbf{1}\right]|\tilde{o}\rangle\\
&=X|\tilde{o}\rangle.
\end{aligned}
\eeq
where the $4N^2\times4N^2$ matrix $X$ is called the damping matrix, with $N$ being the length of the spin chain. Diagonalizing the damping matrix as
\beq
X=\sum_n\lambda_n|nR\rangle\langle nL|,
\eeq
we have
\beq
|\tilde{o}(t)\rangle=\sum_ne^{\lambda_nt}|nR\rangle\langle nL|\tilde{o}(0)\rangle,
\eeq
and this allows to compute $C(t)=\text{tr}(\sigma_N^zO(t))$ as
\beq
C(t)=\langle\tilde{s}_N^z|\tilde{o}(t)\rangle
\eeq
where $|\tilde{s}_N^z\rangle$ is the vectorized form of $s_N^z$, and $s_N^z$ is the representation of $\sigma_N^z$ under the Majorana basis:
\beq
\sigma_N^z=\frac{i}{4}\vec{\gamma}^Ts_N^z\vec{\gamma}.
\eeq
In \figref{figS3}, it is shown the spectrum of the damping matrix $X$ for $N=20$ and $B=0$ and $B=0.02$, respectively. In the non-solvable case $B=0.02$, the spectrum is more complicated than the solvable case $B=0$. The spectrum also indicates a long-time relaxation rate of $\Delta=0.65$ for $B=0$, and $\Delta=0.47$ for $B=0.02$, in agreement with the results in \figref{fig3} in the main text.


\bibliography{main,NH_bib}{}

\end{document}